\documentclass[sigconf]{acmart}

\newcommand{\approach}{{CF$^2$}}
\usepackage{chemformula}
\usepackage{amsmath}
\usepackage{multirow}
\usepackage{booktabs}
\usepackage{amsfonts}
\usepackage{adjustbox}
\DeclareMathOperator*{\argmax}{arg\,max}
\DeclareMathOperator*{\minimize}{minimize}

\AtBeginDocument{%
  \providecommand\BibTeX{{%
    \normalfont B\kern-0.5em{\scshape i\kern-0.25em b}\kern-0.8em\TeX}}}

\copyrightyear{2022}
\acmYear{2022}
\setcopyright{acmlicensed}
\acmConference[WWW '22]{Proceedings of the ACM Web Conference 2022}{April 25--29, 2022}{Virtual Event, Lyon, France}
\acmBooktitle{Proceedings of the ACM Web Conference 2022 (WWW '22), April 25--29, 2022, Virtual Event, Lyon, France}
\acmPrice{15.00}
\acmDOI{10.1145/3485447.3511948}
\acmISBN{978-1-4503-9096-5/22/04}

\begin{document}

\title[Learning and Evaluating Graph Neural Network Explanations]{Learning and Evaluating Graph Neural Network Explanations based on Counterfactual and Factual Reasoning}

\author{Juntao Tan, Shijie Geng, Zuohui Fu, Yingqiang Ge, Shuyuan Xu, Yunqi Li, Yongfeng Zhang}
\affiliation{%
  \institution{Department of Computer Science, Rutgers University, New Brunswick, NJ 08854, US}
     \country{}
  \{juntao.tan, shijie.geng, zuohui.fu, yingqiang.ge, shuyuan.xu, yunqi.li, yongfeng.zhang\}@rutgers.edu
}


\begin{abstract}
    Structural data well exists in Web applications, such as social networks in social media, citation networks in academic websites, and threads data in online forums. Due to the complex topology, it is difficult to process and make use of the rich information within such data. Graph Neural Networks (GNNs) have shown great advantages on learning representations for structural data. However, the non-transparency of the deep learning models makes it non-trivial to explain and interpret the predictions made by GNNs. Meanwhile, it is also a big challenge to evaluate the GNN explanations, since in many cases, the ground-truth explanations are unavailable. 
    
    In this paper, we take insights of \textbf{C}ounter\textbf{f}actual and \textbf{F}actual (\approach) reasoning from causal inference theory, to solve both the learning and evaluation problems in explainable GNNs. For generating explanations, we propose a model-agnostic framework by formulating an optimization problem based on both of the two casual perspectives. This distinguishes \approach~from previous explainable GNNs that only consider one of them. Another contribution of the work is the evaluation of GNN explanations. For quantitatively evaluating the generated explanations without the requirement of ground-truth, we design metrics based on Counterfactual and Factual reasoning to evaluate the \textit{necessity} and \textit{sufficiency} of the explanations. Experiments show that no matter ground-truth explanations are available or not, \approach~generates better explanations than previous state-of-the-art methods on real-world datasets. Moreover, the statistic analysis justifies the correlation between the performance on ground-truth evaluation and our proposed metrics. Source code is available at \url{https://github.com/chrisjtan/gnn_cff}.
\end{abstract}




\maketitle

\section{Introduction}
\begin{figure}
    \centering
    \includegraphics[width=0.95\linewidth]{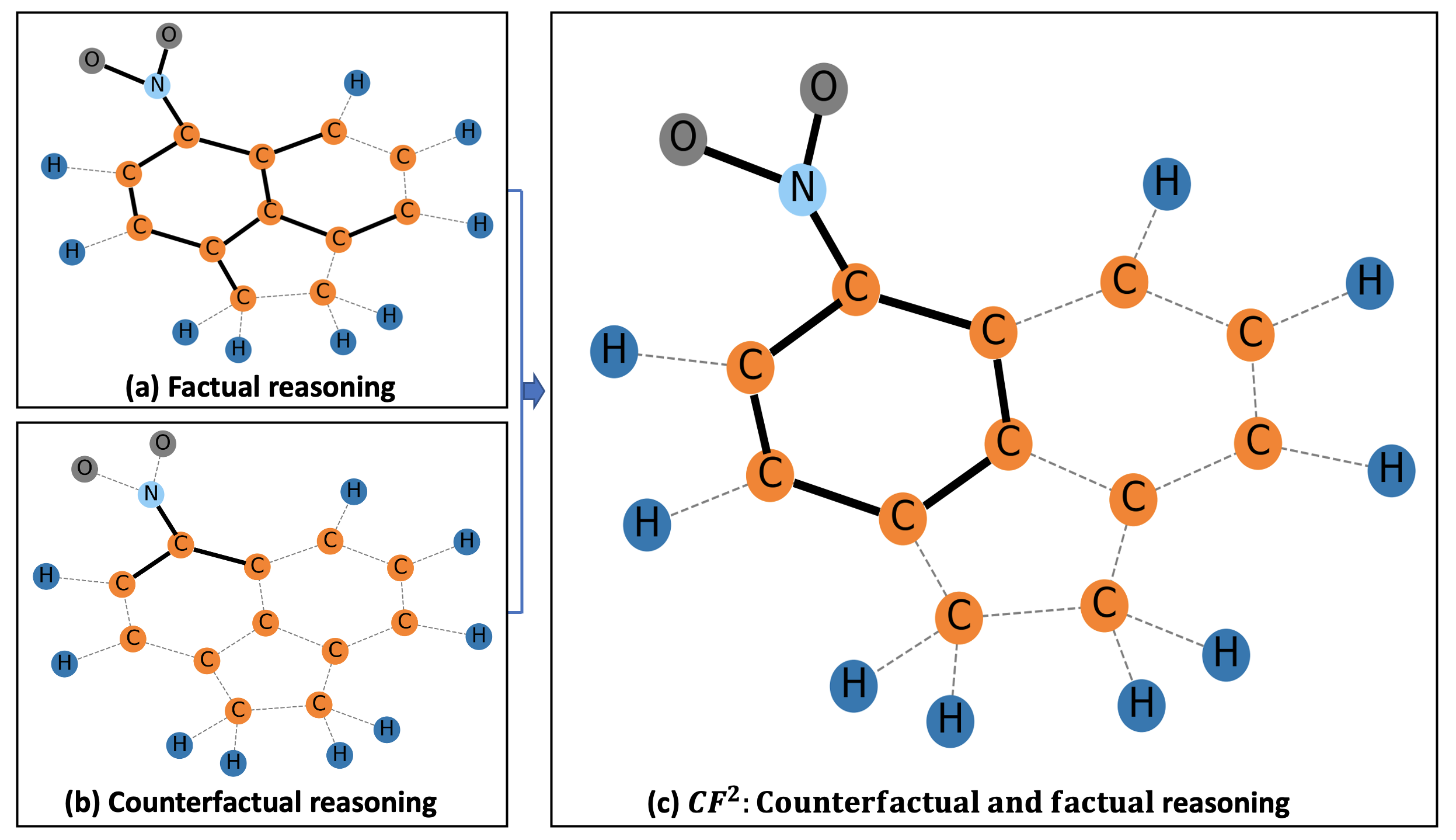}
    \caption{An example for extracting explanations for mutagenic prediction. The sub-graph induced by the bold edges is the explanation extracted by (a) factual reasoning, (b) counterfactual reasoning and (c) counterfactual and factual reasoning. The sub-graph in (c) is also the ground-truth explanation, i.e., Nitrobenzene structure is the cause of mutagen.
    }
    \label{fig:overview}
    \vspace{-10pt}
\end{figure}

Structured data widely exists in various domains such as social networks \cite{Yanardag2015redditbinary}, citation networks \cite{Getoor2005citeseer, Sen2008citeseer} in Web applications, and chemical molecules \cite{Debnathmutag, walenci1} in biomedical research. Such kind of data, which is commonly represented as graph, contains rich information. However, conducting studies on graph data is exhausting for human because both the topology information and the node features 
need to be considered. 

Fortunately, GNNs have shown great advantages on learning graph representations because they aggregate both the feature and structure information by passing the massages in the graph. Thus, GNN-based models achieved promising results in graph prediction tasks such as graph classification, node classification, and link prediction. However, most of the GNN models are non-transparent, which leads to the lack of explainability in model predictions. Exploring the explainability of GNNs is crucial because good explanations not only help to understand the model predictions but also help to identify potential flaws in the model and further refine the GNN model.


In a high-level view, recent state-of-the-art GNN explanation methods are based on either factual reasoning \cite{ying2019gnnexplainer, luo2020parameterized} or counterfactual reasoning \cite{lucic2021cfgnnexplainer, lin2021gem}. Methods based on factual reasoning seek a sub-graph whose information is \textit{sufficient} to produce the same prediction as using the whole original graph, while methods based on counterfactual reasoning seek a sub-graph whose information is \textit{necessary} which if removed will result in different predictions. 

Both factual reasoning and counterfactual reasoning are important approaches to explanation extraction, but each of them alone has its disadvantages. Factual reasoning favors sub-graph explanations that contain enough information to make the same prediction, but the extracted sub-graph may include redundant nodes/edges and thus not compact enough. For example, an extreme case is to take the whole graph as the ``sub-graph,'' which will definitely give the same prediction, but such a ``sub-graph'' does not convey any meaningful information as an explanation. 

This disadvantage is also illustrated in Figure \ref{fig:overview}(a). We use a real-world biochemical example since it has known ground-truth explanation which is hardly accessible for most Web-based graphs. In this example, a molecule is predicted to be mutagenic and we want to extract explanations for the prediction. The explanation sub-graph generated by factual reasoning may indeed cover the essential reason---the Nitrobenzene structure (benzene-\ch{NO2}) \cite{Debnathmutag}. However, it also contains some extra edges from other carbon rings because when these edges are included, the sub-graph leads to the same mutagenic prediction. In a nutshell, the extracted explanation tends to be sufficient but not necessary.

On the other hand, counterfactual reasoning favors the explanations that only contain the most crucial information, i.e., if the explanation sub-graph is removed, then the graph will result in different predictions. However, because of the this, counterfactual reasoning may only extract a small subset of the real explanation.

Take Figure \ref{fig:overview}(b) as an example, counterfactual reasoning generates a sub-graph with only three edges. These edges, if removed, will indeed break the Nitrobenzene structure and thus lead to a different prediction (i.e., non-mutagenic), however, such an explanation does not cover the complete information about what makes the target molecule mutagenic. In a nutshell, the extracted explanation tends to be necessary but not sufficient.

To overcome the problems and to seek a balance between necessity and sufficiency, we propose a Counterfactual and Factual (\approach) reasoning framework to extract GNN explanations which brings the best of the two worlds. \approach ~formulates an optimization problem to integrate counterfactual and factual reasoning objectives so as to
extract explanations that are both necessary and sufficient. As shown in Figure \ref{fig:overview}(c), the counterfactual objective encourages the necessary edges while the factual objective ensures that the extracted explanation contains sufficient information, and thus an ideal sub-graph explanation can be induced.

Another challenge in explainable GNN research is that most real-world graph datasets lack ground-truth explanations, which makes it difficult to evaluate the extracted explanations for these datasets. Fortunately, the fundamental idea of \approach ~can also be adapted into the evaluations. In this paper, we borrow insights from causal inference theory and adopt the \emph{Probability of Necessity} (PN) and \emph{Probability of Sufficiency} (PS) to evaluate the necessity and sufficiency of the extracted explanations, which makes it possible to conduct quantitative evaluation of GNN explanations. PN and PS are aligned with counterfactual and factual reasoning respectively. Details are formulated in Section \ref{sec:metrics}. 


In summary, this work has the following contributions:
\vspace{-3pt}
\begin{itemize}
\setlength\itemsep{0pt}
    \item We show the relationship between factual (or counterfactual) reasoning and the sufficiency (or necessity) of GNN explanations.
    \item We propose a \approach~framework to consider both factual and counterfactual reasoning for GNN explanations.
    \item We propose a set of quantitative evaluation metrics to evaluate the GNN explanations.
    \item We conduct extensive experiments on 2 synthetic datasets and 3 real-world datasets from different domains to justify the proposed model and evaluation metric.
\end{itemize}

\section{Related Works}
\label{sec:related}
\subsection{Explainability in Deep Learning and AI}
Explainable AI has been an important topic in recommender systems \cite{zhang2020explainable,zhang2014explicit,tan2021counterfactual,xian2019reinforcement,chen2021neural,chen2021graph, geng2022path}, natural language processing \cite{hancock2018training, li2021personalized,danilevsky2020survey} and computer vision \cite{escalante2018explainable,vermeire2022explainable, mac2018teaching, chu2020visual, goyal2019counterfactual}. To improve the transparency of deep neural networks, many explanation techniques have been proposed in recent years. Based on how to obtain the importance scores, these approaches can be categorized into gradient/feature-based methods, perturbation/casual-based methods, and surrogate methods \cite{yuan2021explainability, madsen2021post}. Gradients/feature-based methods \cite{simonyan2013deep, li2016understanding, selvaraju2017grad} are the most straightforward way to achieve saliency maps as explanations. They usually map the final prediction to the input space by gradient back-propagation or by linking hidden features to inputs via interpolation. Perturbation/casual-based methods \cite{feng2018pathologies, goyal2019counterfactual, vig2020investigating, tang2020unbiased, niu2021counterfactual, finlayson2021causal, tan2021counterfactual} learn the feature importance through observing the change of predictions with respect to the input perturbation. The idea behind these methods are intuitive: determining which part of the inputs are important by either removing the least important information (i.e., pixels in image, words in text, nodes in graph) to keep the model prediction the same (factual reasoning) or removing the most important information to change the model prediction (counterfactual reasoning). The representative of surrogate methods is LIME \cite{ribeiro2016should}, which employs a simple linear model to approximate the predictions on a bunch of nearby inputs and provides explanations from the surrogate model. 

\subsection{Explainability in Graph Neural Networks}
The aforementioned methods are developed mainly for images and texts. Besides the individual features, graphs also contain important topological structure. Such graph structures are highly related to the functionalities in specific domains and should not be ignored for GNN-based explanation approaches. In explainable GNN, early attempts directly extend gradients/feature-based methods \cite{baldassarre2019explainability, pope2019explainability} to identify important input features. While simple and efficient, these approaches either suffer from gradient saturation \cite{shrikumar2017learning} or lack of the ability to explain node classification predictions \cite{yuan2021explainability}.
Another line of work \cite{huang2020graphlime} follows LIME and adopts a surrogate model for explaining deep graph models. But it ignores the graph structure and cannot explain graph classification models. Hence, these approaches 
are not suitable for explaining the graph-level predictions of GNNs. To solve the problem, \citet{ying2019gnnexplainer} proposed GNNExplainer which treats explanation generation as a mask optimization problem. It follows the idea in perturbation/casual-based methods and learns soft masks that cover the key nodes and edges while maintaining the original prediction score. GISST \cite{lin2020graph} further extended GNNExplainer by identifying important sub-graphs and generating importance scores for all nodes and edges through a self-attention layer. The above two methods learn soft masks that contain continuous values, which suffer from the ``introduce evidence'' problem \cite{yuan2021explainability}. To solve the problem, PGExplainer \cite{luo2020parameterized} adopts the reparameterization trick and learns approximate discrete masks that maximizes the mutual information between key structures and predictions, and XGNN \cite{yuan2020xgnn} generates a graph based on reinforcement learning to approximate the prediction of the original graph. 
As generative models, they also facilitate the holistic explanation for multiple instances. 
Apart from these factual reasoning approaches, there are also recent works exploring counterfactual reasoning. CF-GNNExplainer \cite{lucic2021cfgnnexplainer} introduces counterfactual reasoning to renovate GNNExplainer and is able to generate minimal yet crucial explanations for GNNs. Gem \cite{lin2021gem} distills ground-truth explanations based on Granger causality (a type of counterfactual reasoning) and then trains an auto-encoder architecture to generate adjacency matrix as explanations based on supervised learning. However, these GNN-based explanation approaches only consider factual or counterfactual reasoning alone, and thus will bias towards either sufficiency or necessity rather than achieving a balance when extracting explanations. In this paper, we seek to integrate counterfactual and factual reasoning to extract GNN explanations that are both sufficient and necessary.

\section{Preliminaries and Notations}
In this section, we briefly introduce how GNNs learn the node and graph representations, as well as its application in the node classification and graph classification tasks. We also introduce the basic notations to the used throughout the paper.

\subsection{Learning Representations}
Given a graph $G = \{\mathcal{V}, \mathcal{E}\}$, and each node $v_i \in \mathcal{V}$ has a $d$-dimensional node feature $x_i \in \mathbb{R}^d$. GNN learns the representation of $v_i$ by iteratively aggregating the information of its neighbors $N(i)$. At the $l$-th layer of a GNN model, $v_i$'s representation $h_i=\text{update}(h_i^{l-1}, h_{N(i)}^l)$, where $h_i^{l-1}$ is the representation of $v_i$ in the previous layer, and $h_{N(i)}$ is aggregated from the neighbors of $v_i$ via an aggregation function: $h_{N(i)}=\text{aggregate}(h_j^{l-1}, \forall v_j\in N(i))$. The implantation of the update($\cdot$) and aggregate($\cdot$) functions can be different for different GNN models. For a GNN model with $L$ layers in total, $h_i^L$ is the final representation of the node $v_i$.

After aggregating the node representations, the graph representation can be computed by taking the average of all the node representations in the graph.

\subsection{Graph Classification}
Given a set of $n$ graphs $\mathcal{G}=\{G_1, G_2, \cdots, G_n\}$, and each graph $G_k \in \mathcal{G}$ is associated with a ground-truth class label $y_k \in \mathcal{C}$, where $\mathcal{C}=\{1, 2, \cdots, r\}$ is the set of graph classes.
The graph classification task aims to learn a graph classifier $\Phi$ that predicts the estimated label $\hat{y}_k$ for an input graph $G_k$. 

Each input graph $G_k=\{\mathcal{V}_k,\mathcal{E}_k\}$ is associated with an adjacency matrix $A_k\in \{0, 1\}^{\mid \mathcal{V}_k\mid \times \mid \mathcal{V}_k\mid}$ and a node feature matrix $X_k \in \mathbb{R}^{\mid \mathcal{V}_k\mid \times d}$. After the training process, the GNN model will predict the estimated label $\hat{y}_k$ for $G_k$ by:
\begin{equation}
    \hat{y}_k = \argmax\limits_{c\in \mathcal{C}} P_\Phi(c\mid A_k, X_k)
\end{equation}
where $\Phi$ is the trained GNN model.

\subsection{Node Classification}
For the node classification task, the goal is to predict the class label for each node in a given graph $G=\{\mathcal{V}, \mathcal{E}\}$. Each node $v_i \in \mathcal{V}$ is associated with a ground-truth node label $y_i\in \mathcal{C}$, where $\mathcal{C}=\{1, 2, \cdots, r\}$ is the set of node classes. In node classification task, since only the $L$-hop neighbors of the node $v_i$ will influence $h_i^{L}$, we define the $L$-hop sub-graph of the node $v_i$ as $G_{s(i)}$ which is the computational graph that will be the input of the GNN model. $A_{s(i)}$ and $X_{s(i)}$ are the related adjacency matrix and feature matrix of the computational sub-graph. The trained GNN model will thus predict the estimated label $\hat{y_i}$ for the node $v_i$ as:
\begin{equation}
    \hat{y}_i = \argmax\limits_{c\in C} P_\Phi(c\mid A_{s(i)}, X_{s(i)})
\end{equation}

\section{Problem Formulation}
\label{sec:problem_formulation}
In this section, we first introduce the explainable GNN problem for the classification task. Then, we mathematically define two objectives for extracting explanations and adjust them into the \approach ~framework. The two objectives are 1) an effective explanation should be both \textit{sufficient} and \textit{necessary}, which are reflected by the factual and counterfactual conditions, respectively; and 2) a good explanation should not only be \textit{effective}, but also be \textit{simple}, which is driven by the Occam's Razor Principle \cite{blumer1987occam}. We formulate the Explanation Strength to reflect the effectiveness and formulate the Explanation Complexity to reflect the simpleness. The above two objectives are the foundation of the \approach ~framework for extracting explanations.

We note that in the rest of the paper, all the concepts, examples and mathematical definitions are introduced under the \textbf{graph classification} problem setting and they can be easily generalized to the node classification task. We provide another version for node classification in Appendix \ref{sec:appendix_a}.

\subsection{Explainable Graph Neural Networks}
\label{sec:concepts}

Suppose a graph $G_k=\{\mathcal{V}_k,\mathcal{E}_k\}$ has the predicted label $\hat{y}_k$, following the setup of \citet{ying2019gnnexplainer}, we generate the explanation for this prediction as a sub-graph, which consists of a subset of the edges and a subset of the feature space of the original graph. The sub-graph can be either connected or unconnected. Thus, the goal of \approach ~is to learn an edge mask $M_k \in \{0, 1\}^{\mid \mathcal{V}_k \mid \times \mid \mathcal{V}_k \mid}$ and a feature mask $F_k \in \{0, 1\}^{\mid \mathcal{V}_k \mid \times d}$, which will be applied on the adjacency matrix $A_k\in \{0, 1\}^{\mid \mathcal{V}_k\mid \times \mid \mathcal{V}_k\mid}$ and the node feature matrix $X_k\in \mathbb{R}^{\mid \mathcal{V}_k\mid \times d}$ of the 
original graph $G_k$. After optimization, the sub-graph will be $A_k \odot M_k$ with the sub-features $X_k \odot F_k$, which is the generated explanation for the prediction of graph $G_k$.

\subsection{Counterfactual and Factual Conditions}
As discussed above, an ideal explanation should be both necessary and sufficient. \approach ~achieves this goal by considering both factual and counterfactual reasoning.

Factual and counterfactual reasoning are two opposite but very symmetric ways of reasoning. Factual reasoning asks the question ``Given A already happened, will B happen?'' Counterfactual reasoning, on the contrary, asks ``If A did not happen, will B still happen?'' \cite{factualconditional}. 
Under the context of GNN explanations, factual reasoning generates sub-edges/sub-features that satisfy the condition ``With these sub-edges/sub-features, which is consistent with the fact, the GNN prediction will be the same.'' Counterfactual reasoning generates sub-edges/sub-features that satisfy the condition ``Without these sub-edges/sub-features, which is inconsistent with the fact, the GNN prediction will be different.'' Intuitively, factual reasoning seeks a \textit{sufficient} set of edges/features that produce the same prediction as using the whole graph, while counterfactual reasoning seeks a \textit{necessary} set of edges/features that if removed will lead to different predictions.

In \approach~, both factual and counterfactual reasoning are formulated into the model. The condition for factual reasoning is mathematically formulated as following:
\begin{align}
\begin{aligned}\label{eq:factual_condition}
    & \mathrm{Condition~for~Factual~Reasoning:} \\
    & \argmax\limits_{c\in \mathcal{C}} P_\Phi(c\mid A_k\odot M_k, X_k \odot F_k) = \hat{y}_k
\end{aligned}
\end{align}

Similarly, the condition for counterfactual reasoning is formulated as:
\begin{align}
\begin{aligned}\label{eq:counterfactual_condition}
    & \mathrm{Condition~for~Counterfactual~Reasoning:} \\
    & \argmax\limits_{c\in \mathcal{C}} P_\Phi(c\mid A_k - A_k\odot M_k, X_k - X_k \odot F_k) \neq \hat{y}_k
\end{aligned}
\end{align}

These two conditions will be reflected as objectives for explanation extraction in the loss function, which will be introduced in Section \ref{sec:framework}.

\subsection{Simple and Effective Explanations}
\label{sec:simple_effective}
According to the Occam's Razor Principle \cite{blumer1987occam}, if two explanations are equally  effective, we tend to prefer the simpler one. To achieve this goal, we introduce Explanation Complexity and Explanation Strength for GNN explanations. These two concepts help \approach~to seek simple and effective explanations for GNN predictions.

Explanation complexity $C(M, F)$ measures how complicated the explanation is, which is defined as the number of edges/features used to construct the explanation. Note that $M$ and $F$ are binary matrices indicating which edges and features are included in the sub-graph explanation. As a result, $C(M, F)$ can be defined as the number of $1$'s in $M$ and $F$ matrices, i.e.,
\begin{align}
    C(M, F) = \|M\|_0 + \|F\|_0
\end{align}
However, to make $C(M, F)$ optimizable, we will relax it from 0-norm to 1-norm. We will explain in Section \ref{sec:framework}.

Explanation strength $S(M, F)$ measures how effective the explanation is.
As mentioned above, an effective explanation should be both sufficient and necessary, which is pursued by the factual and counterfactual conditions (Eq.\eqref{eq:factual_condition} and \eqref{eq:counterfactual_condition}). As a result, the explanation strength can be defined as two parts: factual explanation strength $S_f(M, F)$ and counterfactual explanation strength $S_c(M, F)$, both are the larger the better.

The mathematical definition of $S_f(M, F)$ is consistent with the condition for factual reasoning, which is:
\begin{align}
    S_f(M, F) = P_\Phi(\hat{y}_k\mid A_k\odot M_k, X_k\odot F_k)
\end{align}

On the contrary, $S_c(M, F)$ is consistent with the condition for counterfactual reasoning, which is:
\begin{align}
    S_c(M, F) = - P_\Phi(\hat{y}_k\mid A_k - A_k\odot M_k, X_k - X_k \odot F_k)
\end{align}

Explanation complexity and strength will serve as the learning objective and learning constraint in the explanation extraction algorithm, which will also be introduced in Section \ref{sec:framework}. 

In Table \ref{tab:concepts}, we provide an overview of the relationships among the aforementioned concepts.

\begin{table}[t]
\caption{\approach ~generates explanations with two goals: 1) the explanation should be simple, i.e., low in explanation complexity, which means that the generated explanation sub-graph should have a small number of edges and features, which can be achieved by $0$-norm or $1$-norm regularization. 2) the explanation should be effective, i.e., high in explanation strength. An effective explanation should be both sufficient and necessary. Sufficiency can be achieved via factual reasoning and necessity via counterfactual reasoning.}
\vspace{-5pt}
\label{tab:concepts}
\centering
\begin{tabular}{l|l|l|l}
\hline
Objs & \begin{tabular}[c]{@{}l@{}}Simple \\ ($\downarrow$ Complexity)\end{tabular} & \multicolumn{2}{c}{\begin{tabular}[c]{@{}l@{}}Effective\\ ($\uparrow$ Strength)\end{tabular}} \\ \hline
Measure & \# edges,  \# features & Sufficiency & Necessity\\ \hline
Method & Regularization & Factual & Counterfactual\\ \hline
\end{tabular}
\vspace{-10pt}
\end{table}

\section{The \approach ~Framework}
\label{sec:framework}
In this section, we first introduce the \approach ~constrained optimization framework. Then we provide a relaxed version to make the framework optimizable.
\subsection{\approach ~Optimization Problem}
\approach ~is able to generate explanation for any prediction made by a GNN model. As mentioned before, \approach ~aims to find simple (i.e., low complexity) and effective (i.e., high strength) explanations, which can be shown as the following constrained optimization framework:
\begin{equation}
\label{eq:overall eces}
\begin{aligned}
    &\minimize~~\text{Explanation Complexity}\\
    &\text{s.t.,}~~\text{Explanation is Strong Enough}
\end{aligned}
\end{equation}

According to the mathematical definition of explanation complexity and strength in Section \ref{sec:simple_effective}, for a given graph $G_k$ with predicted label $\hat{y}_k$, Eq.\eqref{eq:overall eces} can be rewritten as:
\begin{align}
\label{eq:constraint optimization}
\begin{aligned}
        &\minimize~~ C(M_k, F_k)\\
    &\text{s.t.,}~~ S_f(M_k, F_k) > P_\Phi(\hat{y}_{k, s}\mid A_k\odot M_k, X_k\odot F_k),\\
    &S_c(M_k, F_k) > - P_\Phi(\hat{y}_{k, s}\mid A_k - A_k\odot M_k, X_k - X_k \odot F_k)
\end{aligned}
\end{align}
where $\hat{y}_{k, s}$ is the label other than $\hat{y}_k$ that has the largest probability score predicted by the GNN model. Intuitively, the constraint aims to ensure that when only using the information in the explanation sub-graph, the predicted label $\hat{y}_k$'s probability is higher than any other label and thus the prediction does not change, while if information in the explanation sub-graph is removed, $\hat{y}_k$'s probability will be smaller than at least one other label and thus the prediction will change.


\subsection{Relaxed Optimization}
Directly optimizing Eq.\eqref{eq:constraint optimization} is challenging because both the objective part and the constraint part are not differentiable. As a result, we relax the two parts to make them optimizable.

For the objective part, we relax the masks $M_k$ and $F_k$ to real values, which are $M_k^* \in \mathbb{R}^{\mid \mathcal{V}_k\mid \times \mid \mathcal{V}_k\mid}$ and $F_k^*\in \mathbb{R}^{\mid \mathcal{V}_k \mid \times d}$. Meanwhile, since the $0$-norm in the original equation is also not differentiable, we use $1$-norm to ensure the sparsity of $M_k^*$ and $F_k^*$, which has been proven to be effective in \cite{Candes2005l1norm, Candes2006l1norm}.

For the constraint part, we relax it as pairwise contrastive loss $L_f$ and $L_c$, where
\begin{align}
\begin{aligned}
    L_f = &\mathrm{ReLU}(\gamma+P_\Phi(\hat{y}_{k, s}\mid A_k\odot M_k^*, \mathcal{X}_k\odot F_k^*)&\\
    &-S_f(M_k^*,F_k^*))
\end{aligned}
\end{align}

Similarly,
\begin{align}
\begin{aligned}
    L_c =& \mathrm{ReLU}(\gamma-S_c(M_k^*,F_k^*)\\
    &-P_\Phi(\hat{y}_{k, s}\mid A_k - A_k\odot M_k^*, \mathcal{X}_k - \mathcal{X}_k \odot F_k^*))
\end{aligned}
\end{align}

After relaxation, Eq.\eqref{eq:constraint optimization} becomes optimizable, which is:
\begin{align}
\label{eq:relaxed}
\begin{aligned}
    &\minimize~~ \|M_k^*\|_1 + \|F_k^*\|_1 + \lambda (\alpha L_f + (1-\alpha) L_c)\\
\end{aligned}
\end{align}

When solving the relaxed optimization equation, the margin value $\gamma$ in Eq.(10) and Eq.(11) is set to $0.5$. After the optimization, $0.5$ is also used as the threshold to be applied on the optimized masks to generate explanations (i.e., when the value in the masks $M^*$/$F^*$ is larger than 0.5, we keep the related edge/feature in the generated explanation).

In Eq.\eqref{eq:relaxed}, the hyper-parameter $\lambda$ controls the trade-off between the explanation complexity and the explanation strength. By increasing $\lambda$, the model will focus more on the effectiveness of the generated explanations but less on the complexity, which may result in a bigger sub-graph and feature space. Another hyper-parameter $\alpha$ controls the trade-off between the sufficiency and the necessity of the generated explanation. By increasing (or deceasing) $\alpha$, the generated explanation will focus more on the sufficiency (or necessity). 

\section{Evaluating GNN Explanations}

\label{sec:metrics}
Most of the real-world datasets for graph/node classification do not have ground-truth explanations, which makes the evaluation of GNN explanations a big challenge for the community. As mentioned in section \ref{sec:problem_formulation}, a good explanation should be both sufficient and necessary, which is aligned with the factual and counterfactual condition, respectively. 

In logic and mathematics, necessity and sufficiency are terms used to describe a conditional or implicational relationship between two statements. Suppose we have $S\Rightarrow N$, i.e., if $S$ happens then $N$ will happen, then we say $S$ is a sufficient condition for $N$. Meanwhile, we have the logically equivalent contrapositive $\neg N\Rightarrow\neg S$, i.e., if $N$ does not happen, then $S$ will not happen, as a result, we say $N$ is a necessary condition for $S$. In light of this idea, we adopt the concepts of Probability of Sufficiency (PS) and Probability of Necessity (PN) from causal inference theory \cite[p.112]{pearl2016causal}, which enable us to conduct quantitative evaluation of the GNN explanations.

\subsection{Probability of Sufficiency}
For an explanation A that is generated to explain event B, suppose A happens then B will happen, then A satisfies the factual condition and A is a sufficient explanation. We define PS as the percentage of generated explanations that are \textit{sufficient} for the instance to acheive the same prediction as using the whole graph. In explainable GNN problem, \textbf{Probability of Sufficiency} is defined as:
\begin{align}
\begin{aligned}
        &\text{PS}=\frac{\sum_{G_k \in \mathcal{G}}\text{ps}_k}{|\mathcal{G}|},
    ~\text{where}~\text{ps}_k=
        \begin{cases}
      1,~\text{if $\hat{y}_k^\prime = \hat{y}_k$}\  \\
      0,~\text{else}
    \end{cases}\\
    &\text{where} ~~ \hat{y}_k^\prime = \argmax\limits_{c\in \mathcal{C}} P_\Phi(c\mid A_k\odot M_k, X_k \odot F_k)
\end{aligned}
\end{align}

Intuitively, PS measures the percentage of graphs whose explanation sub-graph alone can keep the GNN prediction unchanged, and thus it is sufficient.

\subsection{Probability of Necessity}
Similarly, suppose A does not happen then B will not happen, we say A satisfies the counterfactual condition and A is a necessary explanation. We define PN as the percentage of generated explanations that are \textit{necessary} for the instance to achieve the same prediction as using the whole graph. In explainable GNN problem, \textbf{Probability of Necessity} is defined as:
\begin{align}
\begin{aligned}
        &\text{PN}=\frac{\sum_{G_k \in \mathcal{G}}\text{pn}_k}{|\mathcal{G}|},
    ~\text{where}~\text{pn}_k=
        \begin{cases}
      1,~\text{if $\hat{y}_k^\prime \neq \hat{y}_k$}\  \\
      0,~\text{else}
    \end{cases}\\
    &\text{where} ~~ \hat{y}_k^\prime = \argmax\limits_{c\in \mathcal{C}} P_\Phi(c\mid A_k - A_k\odot M_k, X_k - X_k \odot F_k)
\end{aligned}
\end{align}

Intuitively, PN measures the percentage of graphs whose explanation sub-graph, if removed, will change the GNN prediction, and thus it is necessary. 

Both PS and PN are the higher the better. Similar to the definition of $F_1$ score, we use $F_{NS}=\frac{2\cdot \text{PN}\cdot \text{PS}}{\text{PN} + \text{PS}}$ to measure the overall performance of a GNN explanation method.

\section{Experiments}

In this section, we first introduce the datasets and the comparison baselines. Then, we report the main experimental results and the analyses. Finally, we conduct experiments to show the influence of factual and counterfactual reasoning, which helps to gain deeper understanding of the key concepts of the paper. We also conduct studies to justify the effectiveness of the PN/PS-based evaluation.

\subsection{Datasets}
We test our algorithm on two synthetic and three real-world datasets. The two synthetic datasets are BA-shapes and Tree-Cycles, which were introduced in \citet{ying2019gnnexplainer}. We follow exactly the same setup when generating these two datasets. The three real-world datasets are Mutag \cite{Debnathmutag}, NCI1 \cite{walenci1} and CiteSeer \cite{Getoor2005citeseer,Sen2008citeseer}. The Mutag dataset contains 4,337 molecules classified into two categories: mutagenic or non-mutagenic. The NCI1 dataset contains 4,110 chemical compounds which are categorized as either positive or negative to cell lung cancer. The CiteSeer dataset contains 3,312 scientific publications classified into six classes, in which the nodes are the papers and the links represent that one paper is cited by another one.

BA-Shapes, Tree-Cycles and CiteSeer are for node classification, while Mutag and NCI1 are for graph classification. BA-Shapes and Tree-Cycles have ground-truth motifs (i.e., ``house'' and ``cycle'' structures) for explaining the classification since they are human-designed. However, NCI1 and CiteSeer do not have such ground-truth motifs. We would like to especially mention the motifs in the Mutag dataset. \citet{luo2020parameterized} assumed that the nitro group (\ch{NO2}) and amino group (NH$_2$) are the true reasons for mutaginicity and filtered out the mutagens that do not contain them. However, according to \citet{Debnathmutag}, which is the work that published the Mutag dataset, NH$_2$ requires microsomal activation to achieve full mutagenic potency and the dataset is limited to studies without such activation. Thus, NH$_2$ has very small influence in the Mutag dataset. This is also mentioned in \citet{lin2021gem}, which shows that the presence of NH$_2$ has very low correlation with the classification result on this dataset. In fact, benzene-\ch{NO2} is the only discriminative motif 
in this dataset.
As a result, we extract a sub-dataset, Mutag$_0$, which only includes those chemical compounds that contain benzene-\ch{NO2} and are mutagenic, or that does not contain benzene-\ch{NO2} and are not mutagenic. The statistics of the Mutag dataset are shown in Table \ref{tab:mutag}. Table \ref{tab:datasets} provides the statistics of all the datasets used.

\subsection{Baselines}
The comparable baselines in this paper should satisfy such conditions: 1) They generate sub-graphs for explanation; 2) They can generate explanations for any graph dataset, with or without prior knowledge, e.g., \citet{luo2020parameterized} requires explicit motif to generate explanations thus could not be applied on NCI1 and CiteSeer, which is the reason why it is not included. The baselines are as follows:

\textbf{GNNExplainer} \cite{ying2019gnnexplainer}: An explanation model base on perturbation. It selects a compact sub-graph while maximizing the mutual information with the whole graph.

\textbf{CF-GNNExplainer} \cite{lucic2021cfgnnexplainer}: An extension of GNNExplainer by generating explanations based on counterfactual reasoning.

\textbf{Gem} \cite{lin2021gem}: A generative explanation model based on Granger causality, it trains auto-encoder to generate explanation sub-graphs.

\subsection{Experimental Setup}
There are two phases in the experiments: 1) Training the base GNN model for classification; and 2) Generating the explanations. 

For the base model, a GCN with three layers is used for all the datasets. The hidden dimensions are 16 for BA-Shapes, Tree-Cycles, Mutag and NCI1, and 32 for CiteSeer. The model for Mutag and NCI1 datasets requires an extra pooling and fully convolution layers for computing the graph embeddings. We apply ReLU activation function after all the layers except for the last layer, which is followed by a Softmax function for classification. The learning rate is $0.001$ during training for all datasets and the ratio between training and test set is $8:2$. In Table \ref{tab:train}, we report the number of training epochs and the accuracy of the base model we used in this paper. We use the same base model for all the baselines to fairly compare the explanation ability. Since the explanation method is model-agnostic, the base model can be any classification model for graphs.

In the explanation phase, GNNExplainer and Gem require a human-selected $K$ value to decide the size of the explanations in their settings. When implementing these two methods, we follow the same setup in Gem: for the synthetic datasets, we set $K$ equal to the size (\#edges) of the ground-truth motifs, and we set $K=15$ (\#edges) for Mutag and NCI1. We run two experiments on the CiteSeer dataset: edge-based explanation ($K=5$) and feature-based explanation ($K=60$).
CF-GNNExplainer and \approach ~do not require prior knowledge about the $K$ value. The size of explanations are automatically decided by the model themselves via optimization. 

For the hyper-parameters in \approach, the $\lambda$ is decided by 
normalizing the 1-norm loss and the pairwise contrastive loss into the same scale, which are $[500, 500, 1000, 20, 100]$ for BA-Shapes, Tree-Cycles, Mutag$_0$, NCI1, and CiteSeer, respectively. For the $\alpha$ value, we set it to be $0.6$ to make factual reasoning slightly leading the optimization. We will conduct ablation study on $\alpha$ in Section \ref{sec:alpha} to show its influence.

We evaluate the explanation methods based on the graphs in the test dataset. 
Since the BA-Shapes, Tree-Cycles and Mutag$_0$ datasets have ground-truth explanations, we report the Accuracy, Precision, Recall and $F_1$ scores of the generated explanations of each method. Besides, for all datasets, we evaluate the explanation model with the PS, PN and $F_{NS}$ metrics introduced in Section \ref{sec:metrics}. Note that we not only generate explanations based on the edges, but also generate explanations on the node features and test them on the CiteSeer dataset, which is not examined in previous works.

\begin{table}[t]
\footnotesize
\caption{Statistics of the Mutag dataset, the molecules with ``*'' are the graphs we used to build the Mutag$_0$ dataset.}
\vspace{-10pt}
\label{tab:mutag}
\centering
\begin{adjustbox}{width=0.95\linewidth}
\begin{tabular}{c|c|c}
\hline
& w/ benzene-\ch{NO2} & w/o benzene-\ch{NO2} \\ \hline
mutagen     & 448*               & 1,953                 \\ \hline
non-mutagen & 83                & 1,853*                 \\ \hline
\end{tabular}
\end{adjustbox}
\vspace{-5pt}
\end{table}

\begin{table}[t]
\caption{Statistics of all datasets. ``\#ave n'' and ``\#ave e'' are the number of nodes/edges per graph. ``\#feat'' is the number of features. In the ``task'' column, ``node'' and ``graph'' indicate the dataset is used for the node classification task or graph classification task, respectively. The check marks in the ``gt'' column means the existence of ground-truth motifs.}
\vspace{-10pt}
\centering
\setlength{\tabcolsep}{3pt}
\begin{adjustbox}{width=\linewidth}
\begin{tabular}{l|c|c|c|c|c|c|c}
\hline
Dataset       & \#graph &\#ave n & \#ave e & \#class & \#feat & task & gt\\ \hline
BA-Shapes     & 1 & 700 & 4100 & 4 & - & node & $\checkmark$      \\
Tree-Cycles & 1 & 871 & 1950 & 2 & - & node & $\checkmark$       \\
Mutag  & 4337 & 30.32 & 30.77 & 2 & 14        & graph             \\
Mutag$_0$  & 2301 & 31.74 & 32.54 & 2 & 14        & graph & $\checkmark$
\\
NCI1   & 4110 & 29.87 & 32.30      & 2  & 37 & graph             \\
CiteSeer    & 1 & 3312 & 4732 & 6 & 3703 & node           \\ 
 \hline
\end{tabular}
\end{adjustbox}
\label{tab:datasets}
\vspace{-5pt}
\end{table}

\begin{table*}[t]
\caption{Explanation evaluation w.r.t ground-truth. Acc, Pr and Re represent Accuracy, Precision and Recall, respectively. Models with $^\dagger$ are the models that fix the size of explanations with pre-defined $K$ values. For the metrics that measure the overall explanation performance (e.g., F$_1$ score), we use bold font to mark the highest scores. For the metrics that only measure partial performance (e.g., precision, recall), we mark the highest scores with underlines.}
\vspace{-10pt}
\label{tab:gt table}
\centering
\begin{adjustbox}{width=0.95\linewidth}
\begin{tabular}{lccccccccccccc}
\toprule
\multicolumn{2}{l}{\multirow{2}{*}{Models}} & \multicolumn{4}{c}{\textbf{BA-Shapes}} & \multicolumn{4}{c}{\textbf{Tree-Cycles}}  & \multicolumn{4}{c}{\textbf{Mutag$_0$}}\\ 
\cmidrule(lr){3-6}\cmidrule(lr){7-10}\cmidrule(lr){11-14} 
\multicolumn{2}{l}{}                        & Acc\%   & Pr\%    & Re\%    & F$_1$\%   & Acc\%    & Pr\%     & Re\%    & F$_1$\% & Acc\%    & Pr\%     & Re\%    & F$_1$\%  \\ 
\cmidrule(lr){1-14}
\multicolumn{2}{l}{GNNExplainer$^\dagger$}            & 95.25       & 60.08   & 60.08       & 60.08      & 92.78     & 68.06    & 68.06      & 68.06   & 96.96 & 59.71 & 85.17 & 68.85   \\ 
\multicolumn{2}{l}{CF-GNNExplainer}                    & 94.39       & 67.19      & 54.11      & 56.79   & 90.27        & \underline{87.40}       & 47.45      & 59.10 &  96.91 & \underline{66.09} & 39.46 & 47.39    \\ 
\multicolumn{2}{l}{Gem$^\dagger$}                    & \textbf{96.97}       & 64.16      & 64.16      & 64.16      & 89.88        & 57.23       & 57.23      & 57.23  & 96.43 & 63.12 & 47.11 & 54.68      \\ 
\multicolumn{2}{l}{\approach}    & 96.37       & \underline{73.15}   & \underline{68.18}      & \textbf{66.61}      & \textbf{93.26}        & 84.92       &    \underline{73.84}   & \textbf{75.69} &  \textbf{97.34} & 65.28 & \underline{88.59} & \textbf{72.56}     \\
\bottomrule

\end{tabular}
\end{adjustbox}
\end{table*}

\begin{table*}
\caption{Explanation evaluation on PN/PS-based metrics. $\#$exp is the size of the generated explanations. Models with $^\dagger$ are the models that fix the size of explanations with pre-defined $K$ values. For the metrics that measure the overall explanation performance (e.g., F$_{NS}$ score), we use bold font to mark the highest scores. For the metrics that only measure partial performance (e.g., PN, PS), we mark the highest scores with underlines.}
\vspace{-10pt}
\label{tab:PNPS table}
\centering
\begin{adjustbox}{width=0.95\linewidth}
\begin{tabular}{lccccccccccccc}
\toprule
\multicolumn{2}{l}{\multirow{2}{*}{Models}} & \multicolumn{4}{c}{\textbf{BA-Shapes}} & \multicolumn{4}{c}{\textbf{Tree-Cycles}} & \multicolumn{4}{c}{\textbf{Mutag$_0$}} \\ 
\cmidrule(lr){3-6}\cmidrule(lr){7-10}\cmidrule(lr){11-14}
\multicolumn{2}{c}{}                        & PN\%    & PS\%    & F$_{NS}$\%    & \#exp    & PN\%    & PS\%    & F$_{NS}$\%    & \#exp & PN\%    & PS\%    & F$_{NS}$\%    & \#exp     \\ 
\cmidrule(lr){1-14}
\multicolumn{2}{l}{GNNExplainer$^\dagger$}               & 72.19 & 45.62 & 55.91 & 6.00 & 100.00 & 59.72 & 74.78 & 6.00 & 71.79       & 97.44      & 82.67      & 15.00     \\ 
\multicolumn{2}{l}{CF-GNNExplainer}               & 75.34 & 41.10 & 53.18 & 5.79 & 100.00 & 31.94 & 48.42 & 3.44 & 96.26       & 7.48      & 13.88      & 7.72        \\ 
\multicolumn{2}{l}{Gem$^\dagger$}                    & 61.36        &   52.27    & 56.45      & 6.00   & 100.00   &  29.89       & 46.02      & 6.00 & 83.01  & 76.42 & 79.58 & 15.00    \\ 
\multicolumn{2}{l}{\approach}   & \underline{76.73} & \underline{68.22} &  \textbf{72.07} & 6.21 & \underline{100.00} & \underline{81.94} & \textbf{90.08} & 5.81 & \underline{97.44}      & \underline{100.00}      & \textbf{98.70}      & 14.95      \\ 
\toprule

\multicolumn{2}{l}{\multirow{2}{*}{Models}} & \multicolumn{4}{c}{\textbf{NCI1}} & \multicolumn{4}{c}{\textbf{CiteSeer (edge)}} & \multicolumn{4}{c}{\textbf{CiteSeer (feature)}} \\ 
\cmidrule(lr){3-6}\cmidrule(lr){7-10}\cmidrule(lr){11-14}
\multicolumn{2}{c}{}                        & PN\%    & PS\%    & F$_{NS}$\%    & \#exp    & PN\%    & PS\%    & F$_{NS}$\%    & \#exp & PN\%    & PS\%    & F$_{NS}$\%    & \#exp    \\ 
\cmidrule(lr){1-14}
\multicolumn{2}{l}{GNNExplainer$^\dagger$}                & 92.13       & 62.16      & 74.24      & 15.00      & 66.67        & 90.05       & 76.61      & 5.00 & 71.64 & \underline{99.50} & 72.79 & 60.00      \\ 
\multicolumn{2}{l}{CF-GNNExplainer}                 & 97.14       & 31.43      & 47.49      & 7.75      & 69.50 & 82.00 & 75.23 & 2.58& 72.14 & 92.54 & 81.07 & 72.91   \\ 
\multicolumn{2}{l}{Gem$^\dagger$}                    & 99.03 & 52.15      & 68.32      & 15.00      & 61.05  & 72.67       & 66.36      & 5.00 & - & - & - & -    \\ 
\multicolumn{2}{l}{\approach}   & \underline{100.00}      & \underline{63.81}      & \textbf{77.91}      & 17.70      & \underline{71.00} & \underline{94.50} & \textbf{81.08} & 3.18 & \underline{74.63} & 95.02 & \textbf{83.60} & 62.73      \\ 
\bottomrule
\end{tabular}
\end{adjustbox}
\end{table*}

\subsection{Quantitative Analysis}
In Table \ref{tab:gt table}, we report the evaluation of the generated explanations with respect to the ground-truth motifs. \approach ~has an overall better performance than all the other baselines according to Accuracy and F$_1$ scores. The only exception is when comparing with Gem on the BA-Shapes dataset with respect to Accuracy, which is lower by 0.62\%. However, since Gem requires the size of the ground-truth motif to select exactly the same size of explanation, which is a strong prior knowledge, this minor difference is considered acceptable. Another observation is that CF-GNNExplainer is higher in Precision and GNNExplainer is higher in Recall when comparing with each other. This justifies our initial motivation about factual and counterfactual reasoning: The factual reasoning focuses on the sufficiency of the explanation, which results in a higher coverage on the ground-truth motifs, while counterfactual reasoning focuses on the necessity, which provides more precise explanations but worse in coverage. As a result, \approach ~is balancing between them and has an overall higher performance in F$_1$.

Then, for all the datasets, we test the generated explanations with the PN, PS, and F$_{NS}$ scores, as shown in Table \ref{tab:PNPS table}. \approach ~performs the best among all the baselines on PN in $100\%$ cases, on PS in $83\%$ cases, and on F$_{NS}$ in $100\%$ cases. 
Moreover, \approach ~has $13.57\%$ average improvement than the best performance of the baselines on F$_{NS}$, which is significant. Similar to the observations in the ground-truth evaluation, we note that the counterfactual-based methods perform better in PN and factual-based methods perform better in PS. This is in line with our previous analysis on the advantages and disadvantages of factual and counterfactual reasoning. Besides, this result also gives us insights about the relationship between Precision/Recall and PN/PS.


\begin{figure*}[t]
    \centering
    \includegraphics[width=0.8\linewidth]{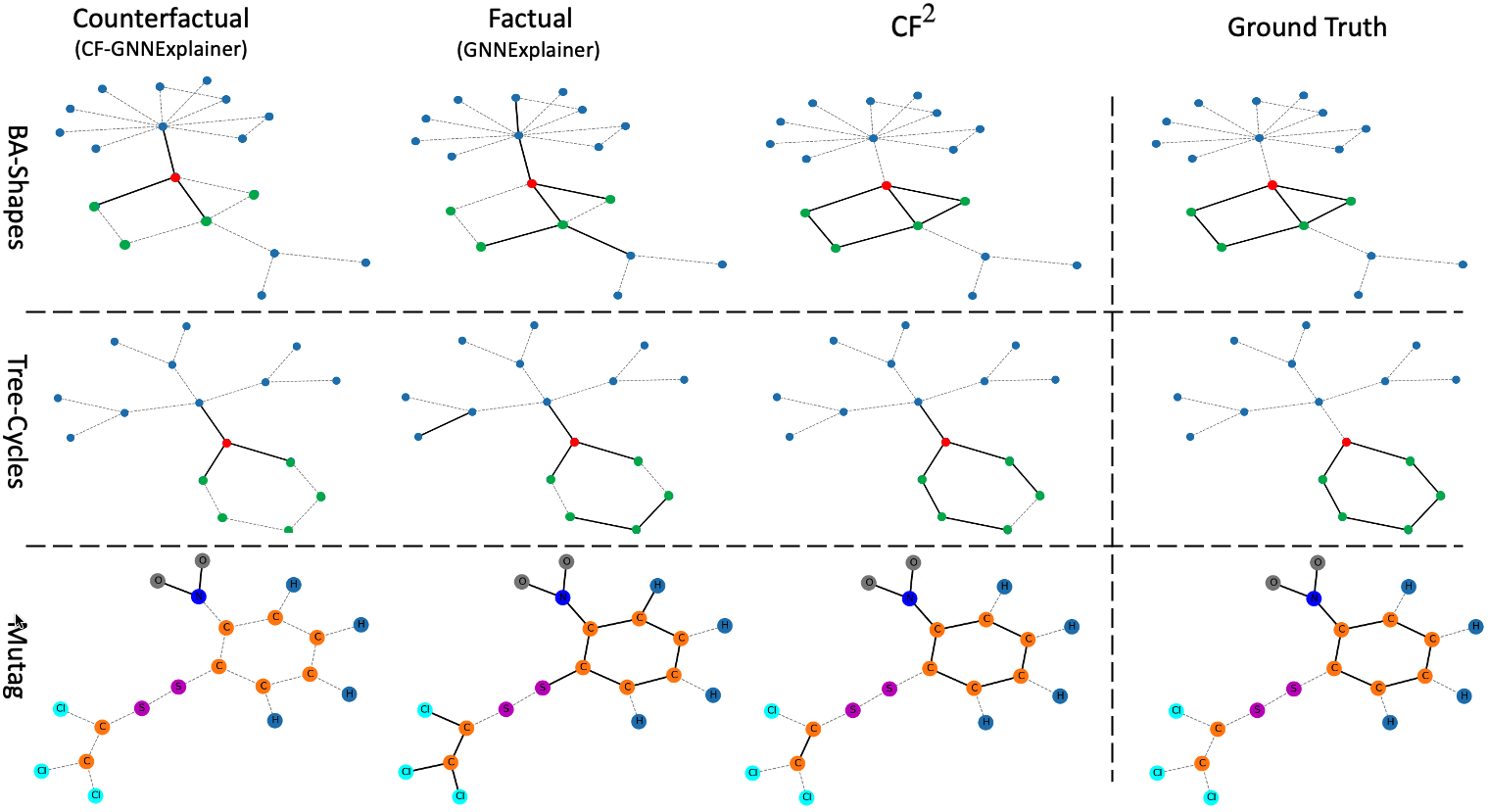}
    \vspace{-10pt}
    \caption{Qualitative Analysis. Illustration of the generated explanations on instances from two synthetic datasets, BA-Shapes and Tree-Cycles, and one real-word dataset, Mutag. From left to right, we show the explanations generated by the methods based on counterfactual reasoning (i.e., CF-GNNExplainer), factual reasoning (i.e., GNNExpainer), \approach, and ground-truth explanation.}
    \label{fig:case}
    \vspace{-5pt}
\end{figure*}

\subsection{Qualitative Analysis}
In Figure \ref{fig:case}, we illustrate explanations based on topology structures to qualitatively compare \approach ~with the methods based on only factual (GNNExplainer) or counterfactual (CF-GNNExplainer) reasoning. Results show that \approach ~better discovers graph motifs than the other two methods. Moreover, counterfactual-based optimization has more precise prediction but tends to be conservative and low in coverage. Factual-based optimization discovers larger portion of the motifs but also covers redundant edges. In general, \approach ~outperforms the other two methods by considering both necessity and sufficiency in the optimization.

\begin{table}[t]
\caption{The classification accuracy of the trained base model on each dataset.}
\label{tab:train}
\centering
\begin{adjustbox}{width=0.95\linewidth}
\begin{tabular}{c|c|c|c|c|c}
\hline
Datasets & BA-Shapes & Tree-Cycles & Mutag$_0$& NCI1 & CiteSeer\\ \hline
Epochs & 3000 & 3000 & 1000 & 200 & 200 \\ \hline
Accuracy     & 97.86  & 98.29 & 98.05 & 69.03 & 71.04                  \\  \hline
\end{tabular}
\end{adjustbox}
\end{table}

\subsection{Influence of $\alpha$}
\label{sec:alpha}

\begin{figure}[t]
    \centering
    \includegraphics[width=0.98\linewidth]{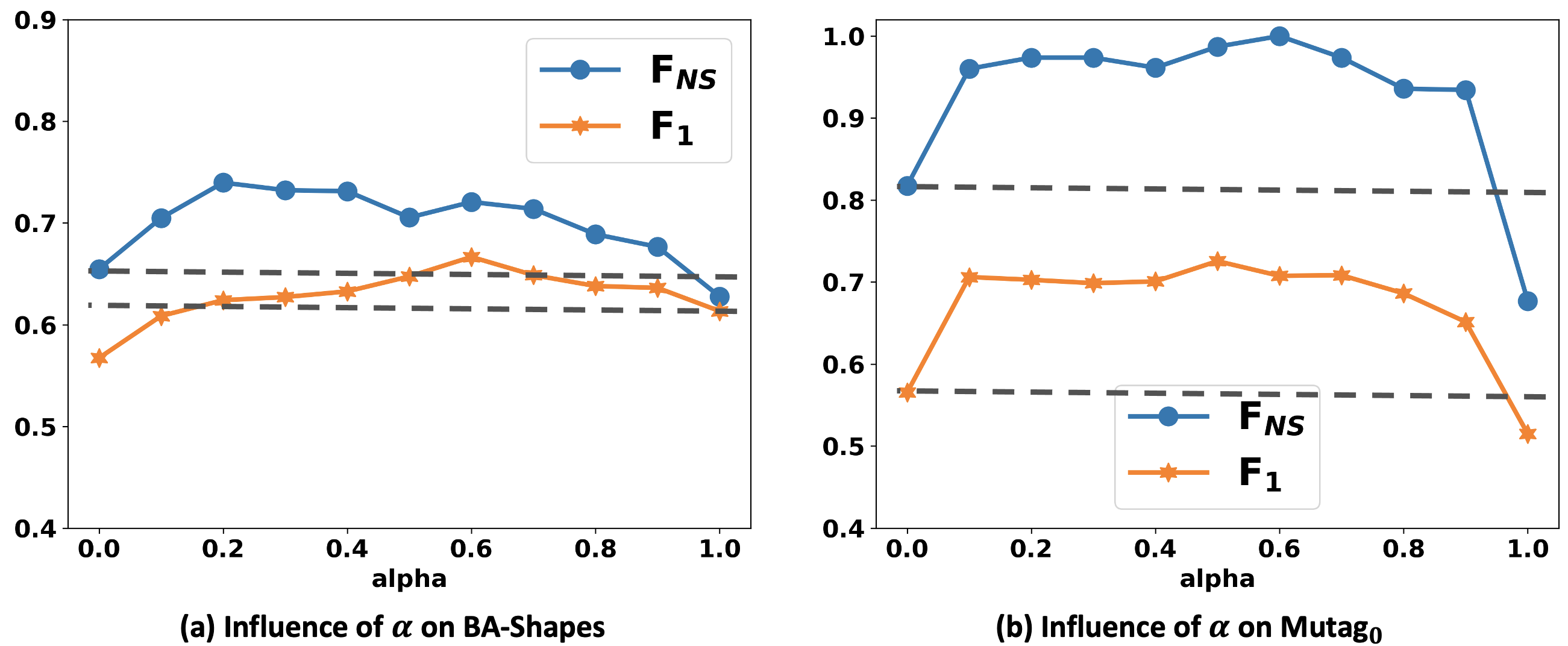}
    \caption{Influence of $\alpha$ on (a) BA-Shapes and (b) Mutag$_0$.}
    \label{fig:alpha}
    \vspace{-15pt}
\end{figure}

The $\alpha$ in Eq.\eqref{eq:relaxed} controls the balance between factual reasoning and counterfactual reasoning. When $\alpha$ is greater than $0.5$, \approach ~considers factual reasoning more than counterfactual reasoning, and when it is less than $0.5$, counterfactual reasoning is considered more than factual reasoning. Figure \ref{fig:alpha} shows the influence of $\alpha$ on \approach ~when generating explanations for BA-Shapes and Mutag$_0$ datasets. Result shows that the value of $\alpha$ is not sensitive, and no matter which $\alpha$ value we choose in $(0, 1)$, the generated explanations are better than only considering one type of reasoning (i.e., $\alpha = 0$ or $\alpha = 1$).

\subsection{Justification of the Evaluation Metric}

\label{sec:justifymetric}

To justify the effectiveness of our PN/PS-based evaluation, we test it on the three datasets with ground-truth explanations, i.e., BA-Shapes, Tree-Cycles, and Mutag$_0$. We use two non-parametric methods to test the correlation between the performance on ground-truth evaluation and PN/PS-based evaluation, which are the Kendall's $\tau$ \cite{kendall1945kendall} and Spearman's $\rho$ \cite{zwillinger1999spearman} scores. These two scores are in the range of (-1, 1). Two values are considered positively correlated if $\tau$ and $\rho$ are positive scores. The higher the scores are, the closer our proposed evaluation metric is compared to ground-truth evaluation, i.e., can be trusted more to evaluate a given explainable GNN model when the ground-truth is not accessible. We test the correlation between F$_{NS}$ and F$_1$/Accuracy. The results are reported in Table \ref{tab:correlation table}. The $\tau$ and $\rho$ show that they are highly positively correlated. This is important since for a dataset without ground-truth motifs, if one explanation method performs better than another one according to the PN/PS-based evaluation, then we can have a good confidence to expect the same conclusion if traditional evaluation metrics are used assuming ground-truth is available.

\section{Conclusions and Future Work}
In this work, we propose a Counterfactual and Factual  reasoning (\approach) framework, which generates GNN explanations by simultaneously considering the necessity and sufficiency of the explanations.
Moreover, we leverage the insights from causal inference theory by taking the Probability of Necessity (PN) and Probability of Sufficiency (PS) to evaluate the necessity and sufficiency of the extracted explanations, making it possible to conduct quantitative evaluation of GNN explanations. 
Experiments on both synthetic and real-world datasets verify the superiority of the proposed method as well as the usefulness of the evaluation metrics. In the future, we will generalize our framework beyond graph-based explanations, including but not limited to vision- and language-based explanations.

\begin{table}[t]
\caption{Correlation between PN/PS-based evaluation and ground-truth evaluation.}
\vspace{-10pt}
\label{tab:correlation table}
\centering
\begin{adjustbox}{width=0.95\linewidth}
\begin{tabular}{lccccccc}
\toprule
\multicolumn{2}{l}{\multirow{2}{*}{Models}} & \multicolumn{2}{c}{\textbf{BA-Shapes}} & \multicolumn{2}{c}{\textbf{Tree-Cycles}}  & \multicolumn{2}{c}{\textbf{Mutag$_0$}}\\ 
\cmidrule(lr){3-4}\cmidrule(lr){5-6}\cmidrule(lr){7-8} 
\multicolumn{2}{l}{}                        & $\tau$ $\uparrow$    & $\rho$ $\uparrow$   & $\tau$ $\uparrow$   & $\rho$  $\uparrow$   & $\tau$ $\uparrow$    & $\rho$ $\uparrow$\\ 
\cmidrule(lr){1-8}
\multicolumn{2}{l}{F$_{NS}$ \& F$_1$} & 1.00 & 1.00 & 1.00 & 1.00 & 1.00 & 1.00  \\ 
\multicolumn{2}{l}{F$_{NS}$ \& Acc} & 0.66 & 0.79 & 1.00 & 1.00 & 0.66 & 0.79  \\ \bottomrule
\end{tabular}
\end{adjustbox}
\vspace{-10pt}
\end{table}

\section*{Acknowledgement}
This work was supported in part by NSF IIS 1910154, 2007907, and 2046457. Any opinions, findings, conclusions or recommendations expressed in this material are those of the authors and do not necessarily reflect those of the sponsors.

\bibliographystyle{ACM-Reference-Format}
\bibliography{main}

\appendix
\newpage
\section{Mathematical Definitions for Node Classification}
\label{sec:appendix_a}
In Section \ref{sec:problem_formulation} and Section \ref{sec:framework}, we formulate the Explainable GNN problem as well as \approach ~framework, under the graph classification setting. In this section, we provide the same mathematical definition under node classification task.
\subsection{Problem Formulation (Node Classification)}
\textbf{Explainable Graph Neural Networks} In a given graph $G=\{\mathcal{V}, \mathcal{E}\}$, Suppose a node $v_i \in G$ has the predicted label $\hat{y}_i$. The computational graph for node $v_i$ is defined as $G_{s(i)}=\{\mathcal{V}_{s(i)}, \mathcal{E}_{s(i)}\}$, which is a sub-graph of $G$ that consists of the $L$-hop neighbors of node $v_i$. $A_{s(i)}\in \{0, 1\}^{\mid \mathcal{V}_{s(i)}\mid \times \mid \mathcal{V}_{s(i)}\mid}$ and $X_{s(i)}\in \mathbb{R}^{\mid \mathcal{V}_{s(i)}\mid \times d}$ are the related adjacency matrix and feature matrix of the computational graph. Since only $G_{s(i)}$ will influence the prediction made by the GNN model, the generated explanation should be a sub-graph of $G_{s(i)}$. Thus, for node classification task, the goal of the explainable GNN problem is to learn an edge mask $M_{s(i)} \in \{0, 1\}^{\mid \mathcal{V}_{s(i)} \mid \times \mid \mathcal{V}_{s(i)} \mid}$ and a feature mask $F_{s(i)} \in \{0, 1\}^{\mid \mathcal{V}_{s(i)}\mid \times d}$, which will be applied on $A_{s(i)}$ and $X_{s(i)}$, respectively. After optimization, the sub-graph will be $A_{s(i)} \odot M_{s(i)}$ with the sub-features $X_{s(i)} \odot F_{s(i)}$, which is the generated explanation for the prediction of node $v_i$.\\
\textbf{Counterfactual and Factual Conditions} For node classification task, the definition of the conditions for factual and counterfactual is similar to graph classification, which are defined as following:
\begin{align}
\begin{aligned}\label{eq:factual_condition_node}
    & \mathrm{Condition~for~Factual~Reasoning:} \\
    & \argmax\limits_{c\in \mathcal{C}} P_\Phi(c\mid A_{s(i)}\odot M_{s(i)}, X_{s(i)} \odot F_{s(i)}) = \hat{y}_i
\end{aligned}
\end{align}

\begin{align}
\begin{aligned}\label{eq:counterfactual_condition_node}
    & \mathrm{Condition~for~Counterfactual~Reasoning:} \\
    & \argmax\limits_{c\in \mathcal{C}} P_\Phi(c\mid A_{s(i)} - A_{s(i)}\odot M_{s(i)}, X_{s(i)} - X_{s(i)} \odot F_{s(i)}) \neq \hat{y}_i
\end{aligned}
\end{align}\\
\textbf{Simple and Effective Explanations} For node classification task, the explanation complexity is defined exactly the same to graph classification, which is:
\begin{align}
    C(M, F) = \|M\|_0 + \|F\|_0
\end{align}
The factual explanation strength and counterfactual explanation strength are defined with the node classification settings as:
\begin{align}
    S_f(M, F) = P_\Phi(\hat{y}_{s(i)}\mid A_{s(i)}\odot M_{s(i)}, X_{s(i)}\odot F_{s(i)})
\end{align}
and
\begin{align}
    S_c(M, F) = - P_\Phi(\hat{y}_{s(i)}\mid A_{s(i)} - A_{s(i)}\odot M_{s(i)}, X_{s(i)} - X_{s(i)} \odot F_{s(i)})
\end{align}
\subsection{The \approach ~Framework (Node Classification)}
The basic idea of \approach ~for node and graph classification are same, which is minimizing the explanation complexity while the generated explanation is strong enough. Therefore, we directly provide the final relaxed optimization and omit the derivation process. \approach ~generates explanations via solving the relaxed optimization equation:
\begin{align}
\label{eq:relaxed_node}
\begin{aligned}
    &\minimize~~ \|M_{s(i)}^*\|_1 + \|F_{s(i)}^*\|_1 + \lambda (\alpha L_f + (1-\alpha) L_c)\\
\end{aligned}
\end{align}
where
\begin{align}
\begin{aligned}
    L_f = &\mathrm{ReLU}(\gamma+P_\Phi(\hat{y}_{i, s}\mid A_{s(i)}\odot M_{s(i)}^*, \mathcal{X}_{s(i)}\odot F_{s(i)}^*)&\\
    &-S_f(M_{s(i)}^*,F_{s(i)}^*))
\end{aligned}
\end{align}

Similarly,
\begin{align}
\begin{aligned}
    L_c =& \mathrm{ReLU}(\gamma-S_c(M_{s(i)}^*,F_{s(i)}^*)\\
    &-P_\Phi(\hat{y}_{i, s}\mid A_{s(i)} - A_{s(i)}\odot M_{s(i)}^*, \mathcal{X}_{s(i)} - \mathcal{X}_{s(i)} \odot F_{s(i)}^*))
\end{aligned}
\end{align}
\end{document}